\documentclass{egpubl}
\usepackage{hpg23}
 
\WsPaper

\usepackage[T1]{fontenc}
\usepackage{dfadobe}  
\usepackage{cite}  

\BibtexOrBiblatex
\electronicVersion
\PrintedOrElectronic
\ifpdf \usepackage[pdftex]{graphicx} \pdfcompresslevel=9
\else \usepackage[dvips]{graphicx} \fi

\usepackage{egweblnk} 
\usepackage{tikz}
\usetikzlibrary{fit,calc,arrows.meta}
\usepackage{pgfplots}
\usepackage{tikz-3dplot}
\usepackage{todonotes}
\usepackage{amsmath}
\usepackage{amssymb}
\usepackage{nicefrac}
\newcommand{\dotprod}[2]{
\left<#1 \cdot #2\right>
}

\title[Real-Time Glints Rendering]%
{Real-Time Rendering of Glinty Appearances \\ using Distributed Binomial Laws on Anisotropic Grids}

\author[T. Deliot \& L. Belcour]
{
\parbox{\textwidth}{\centering T. Deliot\orcid{0000-0001-9675-3442} and L. Belcour\orcid{0000-0002-1982-0717}} \\
{\parbox{\textwidth}{\centering Intel Corporation}}
}

\begin{document}

\teaser{
    \centering
    \vspace{-1.0cm}
    \begin{tikzpicture}
        \node (A) {\includegraphics[width=\linewidth, trim=0 0 0 0, clip]{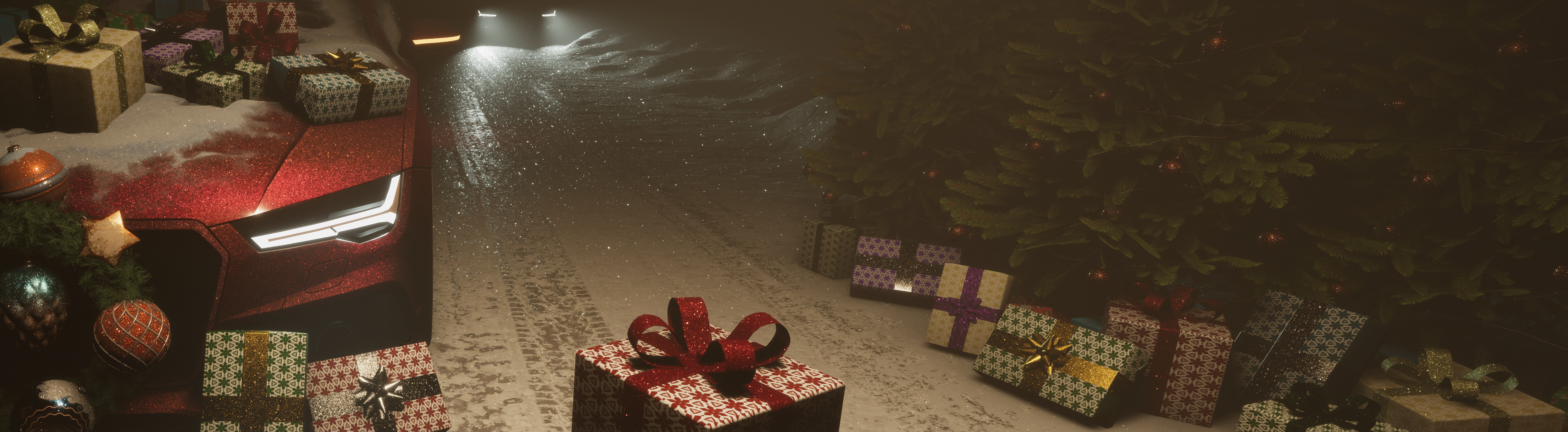}};

        \draw[red, thick] (-3, 1) rectangle +(1, 1);
        \tikzset{shift={(10.15,-1.9)}}
        \begin{scope}
        \clip (-6, 2) rectangle +(2, 2);
        \node (A) {\includegraphics[width=2\linewidth]{figures/fig_teaser/Header_Wide_3840.jpg}};
        \end{scope}
        \draw[red, thick] (-6, 2) rectangle +(2, 2);
        \tikzset{shift={(-10.15,1.9)}}

        \draw[green!70!black, thick] (-6, 0) rectangle +(1, 1);
        \tikzset{shift={(18.5,0.1)}}
        \begin{scope}
        \clip (-12, 0) rectangle +(2, 2);
        \node (A) {\includegraphics[width=2\linewidth]{figures/fig_teaser/Header_Wide_3840.jpg}};
        \end{scope}
        \draw[green!70!black, thick] (-12, 0) rectangle +(2, 2);
        \tikzset{shift={(-18.5,-0.1)}}

        \draw[orange, thick] (-7.8, -1.9) rectangle +(1, 1);
        \tikzset{shift={(19.75,1.6)}}
        \begin{scope}
        \clip (-15.6, -3.8) rectangle +(2, 2);
        \node (A) {\includegraphics[width=2\linewidth]{figures/fig_teaser/Header_Wide_3840.jpg}};
        \end{scope}
        \draw[orange, thick] (-15.6, -3.8) rectangle +(2, 2);
        \tikzset{shift={(-19.75,-1.6)}}

        \draw[blue!50!white, thick] (-5, -2.4) rectangle +(1, 1);
        \tikzset{shift={(16.5,2.6)}}
        \begin{scope}
        \clip (-10, -4.8) rectangle +(2, 2);
        \node (A) {\includegraphics[width=2\linewidth]{figures/fig_teaser/Header_Wide_3840.jpg}};
        \end{scope}
        \draw[blue!50!white, thick] (-10, -4.8) rectangle +(2, 2);
        \tikzset{shift={(-16.5,-2.6)}}
    \end{tikzpicture}
    \vspace{-17pt}
    \caption{We render in real-time glittery appearances such as snow, car paints, or christmass ornaments.}
\label{fig:teaser}
}

\maketitle

\begin{abstract}
In this work, we render in real-time glittery materials caused by discrete flakes on the surface. To achieve this, one has to count the number of flakes reflecting the light towards the camera within every texel covered by a given pixel footprint. To do so, we derive a counting method for arbitrary footprints that, unlike previous work, outputs the correct statistics. We combine this counting method with an anisotropic parameterization of the texture space that reduces the number of texels falling under a pixel footprint. This allows our method to run with both stable performance and $1.5\times$ to $5\times$ faster than the state-of-the-art.

\end{abstract}  

\maketitle
\vspace{-1.5cm}
\section{Introduction}

While rendering glittery appearances greatly enhances the realism of many human-made surfaces (car paints, molded plastics, ...) and natural ones (stones, water, snow, ...), adding it to a rendering engine comes at very high cost. Indeed, such effects require integrating high-frequency details, either due to counting discrete particles~\cite{jakob2014discrete} or integrating high-resolution normal maps~\cite{yan2014rendering}. In contrast to offline rendering, real-time implementations of glittery surfaces must relax different aspect of what defines such models while at the same time preserving their appearance. This is a rather difficult task, and as such, only a handful of methods managed to add this effect to real-time rendering engines~\cite{zirr2016real,chermain2021real}. 

\paragraph*{Specifications for Real-Time.}
One major constraint is that the distribution of details must remain coherent whatever the distance to the camera is. In a nutshell, if a particular glint exists at a given scale, it must remains when the camera zooms out or rotates. Real-time implementations relax this constraint to only \textit{ensure coherence between two consecutive levels-of-detail} (that is, when the camera footprint quadruples its area).

However, real-time methods still require that the underlying stochastic details \textit{converge towards a smooth microfacet model} when the number of details tends to infinity.

\paragraph*{BRDF as a Random Process.}
Using those specifications, Zirr and Kaplanyan~\cite{zirr2016real} derived the fastest glinty shader so far where the \textit{Normal Distribution Function} (NDF) of a microfacet model~\cite{walter2007microfacet} is used as the success probability of multiple coin tosses. By connecting the number of trials to the number of facets within the pixel's footprint, they simulate the randomness of glittery appearances. This process satisfies both constraints as the resulting appearance converges to the continuous microfacet model for an increasing number of trials. They further ensure the coherence between mip-levels by attaching random numbers on the object's surface in texture space.

However, this solution presents two drawbacks: First, it doesn't correctly reproduce the statistics of glittery appearance which \textit{produces artifacts}; Second, the way they attach random numbers to the texture space requires that they manually perform anisotropic filtering. This results in \textit{unstable performance} depending on the viewing conditions and glints density.

\paragraph*{Our solution.}
We use the formulation of Jakob and colleagues~\cite{jakob2014discrete} and search to count the number of discrete facets reflecting light towards the camera.

Similar to Zirr and Kaplanyan~\cite{zirr2016real}, we randomize the evaluation of the NDF by using a stochastic counting process. However, we link this evaluation to an arrangement of facets at the surface of the object and, hence, \textit{ensure a correct evaluation of the statistics of glints}. More precisely, we measure the proportion of flakes within the pixel footprint as if they were randomly distributed over the surface of the object (see Section~\ref{sec:binomials}).

Real-time implementations all rely on discretizing a pixel footprint into overlapping texels at a given level of detail (LOD). When a pixel footprint becomes anisotropic, iterating over all overlapping texels can be a huge computational bottleneck. We show that for glint rendering, we can align the texture parameterizations used for random numbers to pixel footprints and \textit{ensure that, at a given LOD, every footprint maps to a constant number of texels} (see Section~\ref{sec:grids}). This results in both fast and stable performance.

Combined with a \textit{new approximation of the binomial law} and the \textit{use of simplex grids} to fetch random numbers (see Section~\ref{sec:implementation}), we obtain a real-time prototype that offers stable performance and is faster that the state-of-the-art methods of Zirr and Kaplanyan~\cite{zirr2016real} and Chermain et al.~\cite{chermain2020procedural}. We demonstrate that our prototype is $1.5\times $ to $5\times $ faster than theirs (see Section~\ref{sec:results}).

\section{Previous Work}
We split previous work in two categories: offline rendering techniques and real-time methods. We subdivide both categories in two depending on the modeling of the surface geometry: either stochastic particles (glints) or high-frequency normal maps.

\vspace{-5pt}
\subsection{Offline Rendering}

\paragraph*{Stochastic glints.}
This first body of work, pioneered by Jakob et al.~\cite{jakob2014discrete}, represents glints as a discrete set of facets randomly distributed on the surface with orientation following a microfacet model's \textit{Normal Distribution Function} (NDF). The amount of light reflected by the surface depends on the number of reflecting facets in the footprint of the current pixel. This is an expensive 4D hierarchical search.
Atanasov et al.~\cite{atanasov2016practical} improve on Jakob et al.~\cite{jakob2014discrete} model to build a production ready glint appearance model.
Wang et al.~\cite{wang2018fast} accelerate this computation by separating the search of the particle position and orientation. This splits the searching process into two 2D hierarchical searches which is less expensive. They further prefilter the directional statistics to avoid the explicit search in this space.

\paragraph*{High-frequency normal maps}
 While this is not restricted to rendering glittery appearances, another body of work models glittery appearances using high-frequency normal maps~\cite{yan2014rendering}. There, one must integrate over the surface of the pixel footprint, all the normals reflecting light towards the camera.
Chermain et al.~\cite{chermain2020microfacet} use a similar paradigm expressed in the microfacet framework, which permits including multiple scattering effects for example~\cite{chermain2019glint}. Variants of both methods have been explored to reduce the storage or improve the efficiency~\cite{gamboa2018scalable,chermain2020procedural,wang2020example,chermain2021importance,atanasov2021multiscale,deng2022constant}.

\subsection{Real-Time Rendering}
Both the stochastic glints and high-frequency normal maps models have been translated into real-time shaders. In our literature review, we will discard methods not reproducing one of those models such as sparkle shaders~\cite{bowles2015sparkly,wang2016robust}. While they are extremely efficient, they do not offer realistic behaviour or parameters.

\vspace{-3pt}
\paragraph*{Glint models.}
Zirr and Kaplanyan~\cite{zirr2016real} follow the discrete microfacets approach. They replace the explicit counting of the number of reflecting facets by a statistical average using a binomial law. Each facet in the pixel footprint is treated as tossed coin with a probability of reflecting light proportional to the target NDF. Since this greatly simplifies the complexity of the computation, they achieve real-time performance. Still, their method presents artifacts and remains too expensive for video games. Alternatives~\cite{wang2020real} propose different approximations of the counting process but all lead to poorer performance (in the order of $10 ms$). Our aim is to provide a model that is \textit{artifact-free}, has \textit{stable} performance and runs \textit{faster} than existing methods.

\vspace{-3pt}
\paragraph*{Patch based models.}
Inside the formulation of high-frequency normal maps, Chermain et al.~\cite{chermain2020procedural} achieve real-time performance but with physically-based, energy-conserving results. They separate the directional 2D statistics as products of 1D statistics. To avoid explicit counting, they precompute a dictionary of filtered 1D NDFs that account for the accumulation of normals within the pixel footprint. This later part introduces several artifacts and limitations. Notably, they cannot use the GGX NDF. We relax the PBR constraint while preserving the appearance and provide a model that is compatible with any NDF.

\vspace{-3pt}
\paragraph*{Scope of our work.}
In this work, we build on the aggregate counting process of Zirr and Kaplanyan~\cite{zirr2016real}. We show how to maintain the correct counting statistics through levels-of-detail (Section~\ref{sec:binomials}), and how to reduce the number of calls to the counting method within a pixel footprint (Section~\ref{sec:grids}). First, we provide an overview of the underlying shading model (Section~\ref{sec:overview}).

\vspace{-10pt}
\section{Real-Time Rendering of Glints}
\label{sec:overview}

Our shading model starts from the Rendering Equation~\cite{kajiya1986rendering}:
\begin{align}
    L_\mathcal{P} = \int_{\Omega} \rho_\mathcal{P}(\omega_i, \omega_o) \, L_i(\omega_o) \, \mbox{cos}(\theta_o) \, \mbox{d}\omega_o,
\end{align}
averaging over hemisphere $\Omega$ the product of the incoming light $L_i$ with the mean reflectance model over the pixel footprint $\mathcal{P}$:
\begin{align}
   \rho_\mathcal{P}(\omega_i, \omega_o) = {F(\dotprod{\omega_i}{\mathbf{h}}) \, G(\omega_i, \omega_o) \, D_\mathcal{P}(\mathbf{h}) \over 4  \dotprod{\omega_i}{\mathbf{n}} \, \dotprod{\omega_o}{\mathbf{n}}}.
\end{align}
Here, $\mathbf{h}$ is the half-vector between the view $\omega_i$ and light $\omega_o$ vectors, and $F(\cdot)$ and $G(\cdot)$ are respectively the Fresnel and Shadowing/Masking terms.
This model behaves like a smooth microfacet model with the NDF, $D_\mathcal{P}(\mathbf{h})$, replaced by a counting process $c(\cdot)$:
\begin{align}
    D_\mathcal{P}(\mathbf{h}) = {c(N(\mathcal{P}), D(\mathbf{h})) \over N(\mathcal{P})},
\end{align}
with $\mathbb{E}\left[D_\mathcal{P}(\mathbf{h})\right]=D(\mathbf{h})$, and $N(\mathcal{P})$ is the number of discrete microfacets inside the pixel footprint, and $D(\mathbf{h})$ is a standard NDF function such as GGX or Beckmann.
While Jakob et al.~\cite{jakob2014discrete} define an explicit counting procedure of the microfacets, it is too computationally intensive for real-time constraints.

\paragraph*{Counting Microfacets}
Instead, we follow the method of Zirr and Kaplanyan~\cite{zirr2016real} and rely on realizations of the binomial law to count the number of reflecting facets:
\begin{align}
    c(N(\mathcal{P}), D(\mathbf{h})) = \dfrac{D(\mathbf{n})}{R} \times b\left(N(\mathcal{P}), R \times \dfrac{D(\mathbf{h})}{D(\mathbf{n})} \right).
\end{align}
Here, $b(N, p)$ is a sampling of the binomial law for $N$ trials with a probability of success of $p$, $D(\mathbf{n})$ is the NDF at the shading normal and $R \in [0,1]$ is a user defined ratio. While $D(\mathbf{n})$ ensures that the probability of success remain in $[0,1]$, $R$ offers control over the amount of glints locally. In the following, we use $p = R \times {D(\mathbf{h}) \over D(\mathbf{n})}$.

$R$ defines another probability of microfacets being reflecting or non-reflecting for any half-vector. This parameter is \textit{ad-hoc} and we cannot completely link it to a physical quantity. However, in practice, it behaves close to a \emph{microfacet roughness} parameter: low $R$ will reduce the amount of glints and increase their intensity.

\vspace{-5pt}
\paragraph*{Convergence to Smooth Model.}
In this formulation, $D_{\mathcal{P}}$ converges to the smooth NDF when the number of microfacets in the pixel footprint increases. Indeed, when $N$ increases, the ratio $\nicefrac{ b(N, p) }{ N}$ converges to the probability $p$\footnote{since $\mathbb{E}\left[{b(N, p) \over N}\right] = p$, and $\mbox{Var}\left[{b(N, p) \over N}\right] = {p \times (1-p) \over N}$.} and all factors cancel out:
\begin{align}
    \lim_{N \to \infty} D_\mathcal{P}(\mathbf{h}) &=\; \dfrac{D(\mathbf{n})}{R} \times \lim_{N\to\infty} \dfrac{b(N(\mathcal{P}), p)}{N(\mathcal{P})} \\
    &=\; \frac{D(\mathbf{n})}{R} \times \frac{R \times D(\mathbf{h})}{D(\mathbf{n})} \;=\; D(\mathbf{h}).
\end{align}

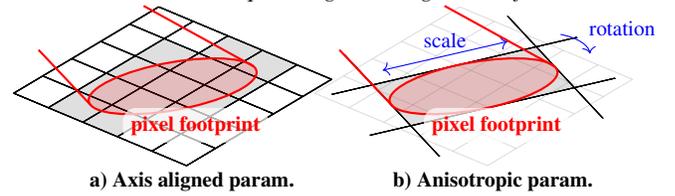
\begin{figure}[t]
    \vspace{-15pt}
    \tdplotsetmaincoords{240}{50}
    \centering
    \begin{tikzpicture}[tdplot_main_coords,font=\footnotesize]
    \begin{scope}[scale=0.6]
    	\begin{scope}[canvas is xy plane at z=0]
            \foreach \x in {-2.5,-1.5,...,2.5}
        		\foreach \y in {-2.5,-1.5,...,2.5}
        		{
        			\draw (\x,-2.5) -- (\x,2.5);
        			\draw (-2.5,\y) -- (2.5,\y);
        		}
        
            \draw[fill=gray,opacity=0.25] ( 1.5,0.5) rectangle +(1.0,1.0);
            \draw[fill=gray,opacity=0.25] ( 0.5,0.5) rectangle +(1.0,1.0);
            \draw[fill=gray,opacity=0.25] (-0.5,0.5) rectangle +(1.0,1.0);
            \draw[fill=gray,opacity=0.25] (-1.5,0.5) rectangle +(1.0,1.0);
            
            \draw[fill=gray,opacity=0.25] ( 1.5,-0.5) rectangle +(1.0,1.0);
            \draw[fill=gray,opacity=0.25] ( 0.5,-0.5) rectangle +(1.0,1.0);
            \draw[fill=gray,opacity=0.25] (-0.5,-0.5) rectangle +(1.0,1.0);
            \draw[fill=gray,opacity=0.25] (-1.5,-0.5) rectangle +(1.0,1.0);
            \draw[fill=gray,opacity=0.25] (-2.5,-0.5) rectangle +(1.0,1.0);

            \draw[fill=gray,opacity=0.25] ( 0.5,-1.5) rectangle +(1.0,1.0);
            \draw[fill=gray,opacity=0.25] (-0.5,-1.5) rectangle +(1.0,1.0);
            \draw[fill=gray,opacity=0.25] (-1.5,-1.5) rectangle +(1.0,1.0);
            \draw[fill=gray,opacity=0.25] (-2.5,-1.5) rectangle +(1.0,1.0);
        
            \draw[rotate=25,color=red,thick,fill=red,fill opacity=0.15] (0,0) ellipse (2cm and 1cm);
        \end{scope}
        \draw[color=red,thick] (-1.0,0,-2.5) -- (1.83,0.8,0);
        \draw[color=red,thick] (-2.0,-2.2,-1.0) -- (-1.83,-0.8,0);
        
        \draw[] (-3.0, 3.0) node {\textbf{ a) Axis aligned param.} };
        \draw[color=red] (-1.0, 1.5) node[fill=white,fill opacity=0.7,text opacity=1.0, rounded corners=5pt] { \textbf{pixel footprint} };
    \end{scope}
    \begin{scope}[xshift=4cm, scale=0.6]
        \begin{scope}[canvas is xy plane at z=0]
        	\foreach \x in {-2.5,-1.5,...,2.5}
        		\foreach \y in {-2.5,-1.5,...,2.5}
        		{
        			\draw[opacity=0.05,color=gray] (\x,-2.5) -- (\x,2.5);
        			\draw[opacity=0.05,color=gray] (-2.5,\y) -- (2.5,\y);
        		}

            \begin{scope}[rotate=25]
                \clip (-3,-2.5) rectangle (3,2.5);  
                \foreach \x in {-6.0,-2.0,...,6.0}
            		\foreach \y in {-3.0,-1.0,...,3.0}
            		{
            			\draw (\x,-4) -- (\x,4);
            			\draw (-6,\y) -- (6,\y);
            		}
                \draw[fill=gray,opacity=0.25] (-2.0,-1.0) rectangle +(4.0,2.0);
            \end{scope}
            
            \draw[rotate=25,color=red,thick,fill=red,fill opacity=0.15] (0,0) ellipse (2cm and 1cm);
            \draw[color=red] (-1.0, 1.5) node[fill=white,fill opacity=0.7,text opacity=1.0, rounded corners=5pt] { \textbf{pixel footprint} };

            \draw[blue,->] (3.0,0) arc (-20:45:1) node[midway,anchor=west,yshift=5pt] {rotation};
            \draw[blue,<->] (-0.75,-2)  -- (2,-0.70) node[midway,anchor=south] {scale};
            
        \end{scope}
        \draw[] (-3.0, 3.0) node {\textbf{ b) Anisotropic param.} };
        \draw[color=red,thick] (-1.0,0,-2.5) -- (1.83,0.8,0);
        \draw[color=red,thick] (-2.0,-2.2,-1.0) -- (-1.83,-0.8,0);
    \end{scope}
    \end{tikzpicture}
    \vspace{-17pt}
    \caption{
        Projected pixel footprints can be highly anisotropic. When the UV space is axis-aligned, the evaluated  number of reflecting glints is distributed over many texels (gray texels - (a)). If we could align the UV-parameterization to the projected pixel footprint, the evaluation would only requires a single texel (b).
        \label{fig:anisotropic_grid}
        \vspace{-20pt}
    }
\end{figure}

\paragraph*{Temporal Coherence}
To ensure that this counting process is temporally stable, random seeds $\theta$ are distributed in a spatial and angular grid and used to evaluate the counting law. Evaluations are bi-linearly interpolated to simulate the spatial and angular life-span of the glint.
In their implementation, Zirr and Kaplanyan~\cite{zirr2016real} also use LOD levels to avoid aliasing. This requires interpolating between two spatial grids at a time to achieve a continuous counting law, which they do linearly. Their binomial law is thus:
\begin{align}
b(N(\mathcal{P}), p) = \mbox{mix}\left(
    b_{\overline\theta}\left(N \times  \overline{\mathcal{P}},  p  \right),
    b_{\underline\theta}\left(N \times \underline{\mathcal{P}}, p  \right),
    w
\right),
\end{align}
where $\mbox{mix}(a, b, w) = (1-w)\times a + w \times b$, and $\overline{\mathcal{P}}$ and $\underline{\mathcal{P}}$ are the grid cell areas of the top and bottom LOD. This interpolation design has two limitations that prevent it from achieving stable real-time performance with no artifacts:

\begin{enumerate}
    \item \textbf{Incorrect Counting.} Interpolating the outputs of the binomial laws between LOD-levels leads to ghosting artifacts (see Figure~\ref{fig:ghosting_binomial} (a)) and produces incorrect statistics (see Figure~\ref{fig:distribute_binomial} (b)). In Section~\ref{sec:binomials}, we detail how interpolating the parameters of the binomials instead of their output solves those issues.
    \item \textbf{Costly Anisotropic Filtering.} When the pixel footprint covers multiple texels, the counting process becomes expensive. Indeed, the counting law needs to be manually evaluated for each texel under the pixel footprint (Figure~\ref{fig:anisotropic_grid} (a)). Because of projection, the pixel footprint can become anisotropic with an arbitrary high spread. This leads to uneven performances since any tilted surface w.r.t. the camera will induce some anisotropy. In Section~\ref{sec:grids}, we provide a parameterization for grids of random numbers that adapts to the anisotropy (see Figure~\ref{fig:anisotropic_grid} (b)).
\end{enumerate}

\input{figures/fig_blending}

\input{figures/fig_distribute_binomial}

\section{Distributed Binomial Laws}
\label{sec:binomials}
The counting method of Zirr and Kaplanyan~\cite{zirr2016real} discards the spatial location of facets within a footprint. However, each facet has a random position at the surface and is a glint if: it is \textit{within the pixel footprint area}; and, it \textit{reflects light}. Both tests are in fact Bernoulli trials with different probabilities. The combined outcome of those tests is what we call a \textit{spatialized Bernoulli trial}. In this work, we assume that both trials are independent and devise a proper interpolation scheme between LOD levels.

\vspace{-5pt}
\subsection{Spatialized Bernoulli Trials}
Let's attach to a random point $\mathbf{x}_i$ on a 2D domain the result of a Bernoulli trial $X_i < p$. We call a spatialized Bernoulli trial the outcome of $\mathbf{x}_i \in \mathcal{A} \cap X_i < p$ for a given area $\mathcal{A}$ and success probability $p$ (see Figure~\ref{fig:distribute_binomial}~(a)). We further define the \textit{spatialized binomial law} that  counts the number of success in $N$ \textit{spatialized} Bernoulli trials. When we use such a counting law to describe a glittery surface, we implicitly take into account the spatial distribution of facets. However, we must to define the spatial distribution of points.

\paragraph*{Uniform Distribution.} If we randomly distribute the trials at the surface of the object, the probability of having a trial within a given area $\mathcal{A}$ is a Bernoulli trial with $p = \mathcal{A}$. It follows that the number of facets within a pixel footprint of area $\mathcal{P}$ becomes:
\begin{align}
    N(\mathcal{P}) =  b(N, \mathcal{P}).
\end{align}
\paragraph*{Stratified Distribution.} When the trials are stratified at the surface of the object (that is, randomly jittered inside a regular grid), the average number of trials for a given area $\mathcal{A}$ is exactly $N_\mathcal{A} =  N \times \mathcal{A}$. Thus, the number of facets within an area $\mathcal{P}$ becomes:
\begin{align}
    N(\mathcal{P}) = \lfloor N \times \mathcal{P} \rfloor.
\end{align}
\paragraph*{Uniform or Stratified?} We described two ways of distributing the Bernoulli trials at the surface of the object. Both relate to ways of modeling glints: a discrete set of random facets~\cite{jakob2014discrete}, or a square texture with exactly one random facet per texel (akin to normal maps~\cite{yan2014rendering}). We opted for the later as the associated counting law was less sensitive to numerical precision.
\vspace{-10pt}

\subsection{Blending of Distributed Binomial Laws}
To express the counting law between LOD levels, we aim at reproducing the statistics of the stratified distribution of Bernoulli trials for intermediate footprints between the LOD levels. 
For a given footprint area $\mathcal{P}$ between a top LOD level of area $\overline{\mathcal{P}}$ and seed $\overline{\theta}$, and bottom LOD level of area $\underline{\mathcal{P}}$ and seed $\underline{\theta}$, we evaluate the number of glints as:
\begin{align}
    b(N \times \mathcal{P}, p) =& \; b_{\overline{\theta}}\left(\overline{w} \times N \times \overline{\mathcal{P}}, p\right) +  b_{\underline{\theta}}\left(\underline{w} \times N \times \underline{\mathcal{P}}, p\right),
    \label{eq:binomial_distribution}
\end{align}
where $\overline{w} = \nicefrac{\mathcal{P} - \underline{\mathcal{P}}}{ \overline{\mathcal{P}} - \underline{\mathcal{P}} }$ and $\underline{w} = \nicefrac{\overline{\mathcal{P}} - \mathcal{P}}{ \overline{\mathcal{P}} - \underline{\mathcal{P}} }$. Intuitively, this blending subdivides the area $\mathcal{P}$ in two: $\mathcal{P} = \overline{w}\times\overline{\mathcal{P}} + \underline{w}\times\underline{\mathcal{P}}$, with $\underline{w} = 1, \overline{w} = 0$ when $\mathcal{P} = \underline{\mathcal{P}}$, and $\overline{w} = 1, \underline{w} = 0$ when $\mathcal{P} = \overline{\mathcal{P}}$. Then, the binomial for the whole area $\mathcal{P}$ is \textit{distributed} as the sum of binomials in each sub-area. This is possible thanks to the additivity of the binomial law: $b(N_1, p) + b(N_2, p) = b(N_1+N_2, p)$. Different distributions are possible (for example distributing over more than 2 areas) but follow the same procedure.

This scheme provides the correct statistics for stratified grids of Bernoulli trials (see Figure~\ref{fig:distribute_binomial}) and avoids ghosting artifacts (see Figure~\ref{fig:ghosting_binomial}). While this corrects the glints' statistics, it still requires a manual anisotropic filtering. In the following Section, we devise a method to always bring the evaluation back to the isotropic case.
\vspace{-10pt}

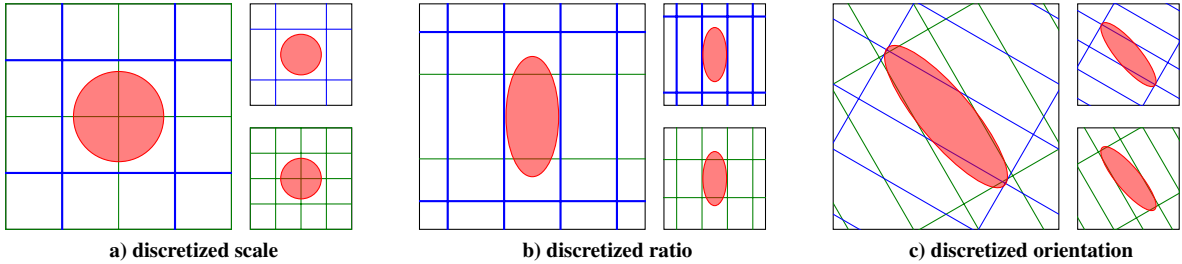
\begin{figure*}[ht]
    \tdplotsetmaincoords{240}{50}
    \centering
    \begin{tikzpicture}[font=\footnotesize]
    \begin{scope}
        \begin{scope}[]
    	\begin{scope}[]
            \draw (0,0) rectangle (3,3);
            \clip (0,0) rectangle (3,3);
            \draw[step=3.0/4.0, shift={(-0.75cm,-0.75cm)}, color=green!50!black]  (0,0) grid (6,6);
            \draw[step=3.0/2.0, shift={(-0.75cm,-0.75cm)}, thick, color=blue] (0,0) grid (6,6);

            \draw[color=red, fill=red, fill opacity=0.5] (1.5, 1.5) ellipse (0.6 and 0.6);
        \end{scope}
        \begin{scope}[xshift=3.25cm, scale=0.45]
            \draw (0,0) rectangle (3,3);
            \clip (0,0) rectangle (3,3);
            \draw[step=3.0/4.0, shift={(-0.75cm,-0.75cm)}, color=green!50!black]  (0,0) grid (6,6);
            
            \draw[color=red, fill=red, fill opacity=0.5] (1.5, 1.5) ellipse (0.6 and 0.6);
        \end{scope}
        
        \begin{scope}[xshift=3.25cm, yshift=1.65cm, scale=0.45]
            \draw (0,0) rectangle (3,3);
            \clip (0,0) rectangle (3,3);
            \draw[step=3.0/2.0, shift={(-0.75cm,-0.75cm)}, color=blue] (0,0) grid (6,6);
            
            \draw[color=red, fill=red, fill opacity=0.5] (1.5, 1.5) ellipse (0.6 and 0.6);
            
        \end{scope}
        
        \node (A) at (2.5,-0.3) {\textbf{a) discretized scale}};
        \end{scope}

        \begin{scope}[xshift=5.5cm]
        \begin{scope}[]
            \draw (0,0) rectangle (3,3);
            \clip (0,0) rectangle (3,3);
            \node (c) at (0,0) {};
            \draw[step=3.0/4.0, yscale=1.5, shift={(-3.0/8.0,-1.0/8.0)}, color=green!50!black]  (0,0) grid (6,6);
            \draw[step=3.0/4.0, yscale=3.0, shift={(-3.0/8.0,-5.0/8.0)}, thick, color=blue]  (0,0) grid (6,6);

            \draw[color=red, fill=red, fill opacity=0.5] (1.5, 1.5) ellipse (0.35 and 0.8);
        \end{scope}
        \begin{scope}[xshift=3.25cm, scale=0.45]
            \draw (0,0) rectangle (3,3);
            \clip (0,0) rectangle (3,3);
            \draw[step=3.0/4.0, yscale=1.5, shift={(-3.0/8.0,-1.0/8.0)}, color=green!50!black]  (0,0) grid (6,6);
            
            \draw[color=red, fill=red, fill opacity=0.5] (1.5, 1.5) ellipse (0.35 and 0.8);
        \end{scope}
        \begin{scope}[xshift=3.25cm, yshift=1.65cm, scale=0.45]
            \draw (0,0) rectangle (3,3);
            \clip (0,0) rectangle (3,3);
            \draw[step=3.0/4.0, yscale=3.0, shift={(-3.0/8.0,-5.0/8.0)}, thick, color=blue]  (0,0) grid (6,6);
        
            \draw[color=red, fill=red, fill opacity=0.5] (1.5, 1.5) ellipse (0.35 and 0.8);
        \end{scope}
        \node (B) at (2.5,-0.3) {\textbf{b) discretized ratio}};
        \end{scope}

        \begin{scope}[xshift=11.0cm]
        \begin{scope}[]
            \draw (0,0) rectangle (3,3);
            \clip (0,0) rectangle (3,3);
            \node (c) at (0,0) {};
            \begin{scope}[shift={(-3/8+1.5, -9/8+1.5)}]
            \begin{scope}[rotate around={30:(3.0/8.0, 9.0/8.0)}]
                \draw[step=3.0/4.0, yscale=3.0, color=green!50!black]  (-3,-3) grid (3,3);    
            \end{scope}
            \end{scope}
            \begin{scope}[shift={(-3/8+1.5, -9/8+1.5)}]
            \begin{scope}[rotate around={60:(3.0/8.0, 9.0/8.0)}]
                \draw[step=3.0/4.0, yscale=3.0, color=blue]  (-3,-3) grid (3,3);    
            \end{scope}
            \end{scope}
            \draw[color=red, fill=red, fill opacity=0.5, rotate around={40:(1.5, 1.5)}, shift={(0,0)}] (1.5, 1.5) ellipse (0.35 and 1.2);
        \end{scope}
        \begin{scope}[xshift=3.25cm, scale=0.45]
            \draw (0,0) rectangle (3,3);
            \clip (0,0) rectangle (3,3);
            \begin{scope}[shift={(-3/8+1.5, -9/8+1.5)}]
            \begin{scope}[rotate around={30:(3.0/8.0, 9.0/8.0)}]
                \draw[step=3.0/4.0, yscale=3.0, color=green!50!black]  (-3,-3) grid (3,3);    
            \end{scope}
            \end{scope}
            
            \draw[color=red, fill=red, fill opacity=0.5, rotate around={40:(1.5, 1.5)}, shift={(0,0)}] (1.5, 1.5) ellipse (0.35 and 1.2);
        \end{scope}
        \begin{scope}[xshift=3.25cm, yshift=1.65cm, scale=0.45]
            \draw (0,0) rectangle (3,3);
            \clip (0,0) rectangle (3,3);
            \begin{scope}[shift={(-3/8+1.5, -9/8+1.5)}]
            \begin{scope}[rotate around={60:(3.0/8.0, 9.0/8.0)}]
                \draw[step=3.0/4.0, yscale=3.0, color=blue]  (-3,-3) grid (3,3);    
            \end{scope}
            \end{scope}
            
            \draw[color=red, fill=red, fill opacity=0.5, rotate around={40:(1.5, 1.5)}, shift={(0,0)}] (1.5, 1.5) ellipse (0.35 and 1.2);
        \end{scope}
        
        \node (C) at (2.5,-0.3) {\textbf{c) discretized orientation}};
        \end{scope}
        
    \end{scope}
    \end{tikzpicture}
    \vspace{-5pt}
    \caption{
        We attach random numbers on a discrete set of grids. For a given pixel filter (in red), we must select the closest grids that encompass the pixel filter's area (grids in blue and green). Because previous works only account for uniform grids with power of two ratios (a), the evaluation of a grid cell can only account for an isotropic footprint. We add the support of anisotropic footprints using grids with a discrete set of ratios (b) and a discrete set of orientations (c).
        Note that we aligned the pixel filter with the grids for visualization purposes.
        \label{fig:selecting_grid}
    }
    \vspace{-15pt}
\end{figure*}

\section{Anisotropic Virtual Grids}
\label{sec:grids}
In our method, similar to previous works~\cite{zirr2016real,chermain2020procedural}, multiple virtual grids of random numbers exist at the surface of the object. For each pixel, the grids with texels best matching the pixel footprint are used to evaluate the counting law. In Zirr and Kaplanyan~\cite{zirr2016real} the grids follow a traditional mip-mapping parameterization: uniform grids of square texels (see Figure~\ref{fig:selecting_grid}~(a)). This can be done by uniformly scaling UV coordinates and keeping equally sized grids:
\begin{align}
    \theta(\mathbf{uv}^\prime) \quad \mbox{with} \; \mathbf{uv}^\prime = \mathcal{S}_{s} \left( \mathbf{uv} \right),
\end{align}
where $\mathcal{S}_{s}$ is the isotropic scale. Since texels are square, distributing the binomial law on such grid approximates the counting of an isotropic filter. Counting anisotropic filter requires accumulating the results of counting in all the texels covered by the filter. We extend the idea of changing the UV parameterization to include non-axis-aligned and non-isotropic grids: by covering the object with grids that are linear transformations of the isotropic grid, we can emulate anisotropic filters (see Figure~\ref{fig:selecting_grid}~(b-c)).

\subsection{Anisotropic Parameterizations}
To ensure that each pixel filter corresponds to one texel, we can transform the UV coordinates using a rotation and a ratio to align the texels with the projected pixel footprint:
\begin{align}
    \mathbf{uv}^\prime = \mathcal{R}_\phi\left( \mathcal{S}_{1, r}\left( \mathcal{S}_{s}\left( \mathbf{uv} \right) \right) \right)
\end{align}
where $\mathcal{S}_{1, r}$, and $\mathcal{R}_\phi$ are respectively the isotropic scale $s$ and anisotropic scale $r$ for the $v$ coordinates, and the rotation of angle $\phi$. In practice, we need to use a discrete set of scales and rotations, which produce a countable set of grids. To find the closest fitting grids for a given pixel filter, we must evaluate its rotation and relative scales.

We start by projecting the pixel footprint on the surface of the object and extract its associated ellipse's major and minor axis $\mathbf{v}^0, \mathbf{v}^1$ in UV space. From this vector, we obtain a rotation $\phi = \mbox{atan}(\mathbf{v}^0_x, \mathbf{v}^0_y)$ and a ratio $r = \nicefrac{|| \mathbf{v}^0 || }{ || \mathbf{v}^1 ||}$. In our implementation, we use screen-space derivatives to obtain the minor/major axis.

\subsection{Interpolating Different Parameterizations}
\label{sec:grids_blending}
In the case of the isotropic parameterization, every pixel footprint's area is enclosed by the cell areas of the isotropic grids:
\begin{align}
    \underline{\mathcal{P}} = \mbox{pow}\left(2, \lfloor\mbox{log}_2\left( \mathcal{P} \right) \rfloor \right), \; \mbox{and}\;
    \overline{\mathcal{P}} = \mbox{pow}\left(2, \lceil \mbox{log}_2\left( \mathcal{P} \right) \rceil  \right)
\end{align}
Similarly, the ratio and orientation of a pixel footprint are enclosed by discrete ratios and angles. This means that $2^3 = 8$ different texels enclose the pixel footprint. We use the blending formula for the distributed binomial law, Equation~\ref{eq:binomial_distribution}, over those $8$ areas. To do so, we assign to each cell a number of facets $N_i$ proportional to its area. We use trilinear weights $w_i$ to distribute the pixel footprint area over each cell's area. We obtain the final distributed binomial value by summing over all cells: $b(N, p) = \sum_{i} b_{\theta_i}\left({w_i \times N_i, p}\right)$.

\begin{figure}[ht]
    \centering
    \vspace{-2pt}
    \begin{tikzpicture}[font=\footnotesize]
    \begin{scope}
        \clip (0, 0) rectangle (5,4.2);
        
        \draw[color=blue] (0, 0) circle (1);
        \draw[color=blue] (0, 0) circle (2);

        \draw[color=blue, rotate=-45.0] (0, 1) -- (0, 4);
        
        \draw[color=blue] (0, 0) circle (3);
        \draw[color=blue, rotate=-22.5] (0, 2) -- (0, 4);
        \draw[color=blue, rotate=-67.5] (0, 2) -- (0, 4);
        
        \draw[color=blue] (0, 0) circle (4);
        \draw[color=blue, rotate=-11.25] (0, 3) -- (0, 4);
        \draw[color=blue, rotate=-33.75] (0, 3) -- (0, 4);
        \draw[color=blue, rotate=-56.25] (0, 3) -- (0, 4);
        \draw[color=blue, rotate=-78.75] (0, 3) -- (0, 4);
    \end{scope}
    \draw[color=blue,->] (0, 0) -- (0.0,4.5);
    \draw[color=blue,->] (0, 0) -- (4.5,0.0);

    \begin{scope}[xshift=-1.75cm, scale=0.45]
        \draw (0,0) rectangle (3,3);
        \clip (0,0) rectangle (3,3);
        \draw[step=3.0/4.0, shift={(-0.75cm,-0.75cm)}, color=green!50!black]  (0,0) grid (6,6);
        
    \end{scope}
    \draw[- Circle] (-0.4, 0.4) -- (-0.2,0.4) -- (-0.2,0.0) -- (0.07, 0.0);

    \begin{scope}[xshift=-1.75cm, yshift=1.60cm, scale=0.45]
        \draw (0,0) rectangle (3,3);
        \clip (0,0) rectangle (3,3);
        \draw[step=3.0/4.0, yscale=1.8, shift={(-3.0/8.0,-3.0/8.0)}, color=green!50!black]  (0,0) grid (6,6);
        
    \end{scope}
    \draw[- Circle] (-0.4, 2.4) -- (-0.2,2.4) -- (-0.2,2.0) -- (0.07, 2.0);

    \begin{scope}[xshift=-1.75cm, yshift=3.2cm, scale=0.45]
        \draw (0,0) rectangle (3,3);
        \clip (0,0) rectangle (3,3);
        \begin{scope}[shift={(-3/8+1.5, -9/8+1.5)}]
            \begin{scope}[rotate around={-30:(3.0/8.0, 9.0/8.0)}]
                \draw[step=3.0/4.0, yscale=3.0, color=green!50!black]  (-3,-3) grid (3,3);    
            \end{scope}
        \end{scope}

    \end{scope}
    \draw[- Circle] (-0.4, 4.2) -- (2.2, 4.2) -- (2.22, 3.25);

    \draw[color=blue, rotate=-45, ->] (0, 4.5) -- node[xshift=-0.2cm, yshift=5pt, rotate=45]{$r$ increase} (0, 6);

    \begin{scope}[xshift=3.5cm, yshift=3cm, rotate=15]
        \draw[blue,->] (0,0) arc (25:-5:5) node[midway,anchor=west] {$\phi$ increase};
    \end{scope}

    \begin{scope}[rotate=-45]
        \node (A00) at (0.0,2.0)[orange, circle, fill, inner sep=1.5]{};
        \node (A01) at (0.0,3.0)[orange, circle, fill, inner sep=1.5]{};
    \end{scope}
    \begin{scope}[rotate=-67.5]
        \node (A10) at (0.0,2.0)[orange, circle, fill, inner sep=1.5]{};
        \node (A12) at (0.0,3.0)[orange, circle, fill, inner sep=1.5]{};
    \end{scope}
    \begin{scope}[rotate=-56.25]
        \node (A11) at (0.0,3.0)[orange, circle, fill, inner sep=1.5]{};
    \end{scope}
    \begin{scope}[rotate=-53.25]
        \node (P) at (0.0,2.42)[red, circle, fill, inner sep=1.5]{};
    \end{scope}
    \draw[->, opacity=0.5] (P) -- (A00);
    \draw[->, opacity=0.5] (P) -- (A01);
    \draw[->, opacity=0.5] (P) -- (A10);
    \draw[->, opacity=0.5] (P) -- (A11);
    \draw[->, opacity=0.5] (P) -- (A12);

    \end{tikzpicture}
    \vspace{-5pt}
    \caption{
        We discretize the change in orientation with respect to the anisotropy ratio: when no anisotropy is required, a single isotopic grid accounts for orientation of the pixel filter. We add one orientation per level of anisotropy ratio. Thus, a pixel filter (red) distributes its binomial law on 5 grids (orange).
        \label{fig:grids_discretization}
    }
\end{figure}
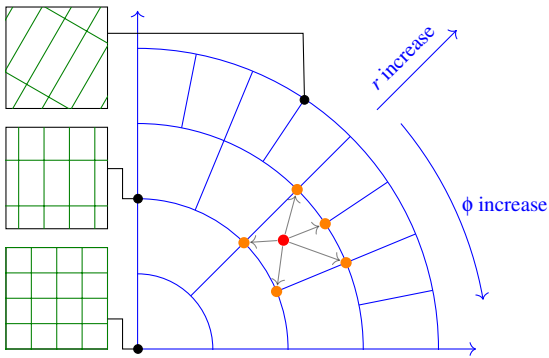

We adapt the discretization of the orientation $\phi$ with respect to the anisotropy ratio $r$. This is to account for the fact that, for isotropic parameterizations ($r=1$) the orientation of the filter should not alter the random numbers $\theta$. We show in Figure~\ref{fig:grids_discretization} the discretization of the orientation with respect to the anisotropy ratio.

\paragraph*{Angular and Spatial Interpolation.}
Virtual grids account for levels-of-detail filtering, to prevent aliasing. In-between these, it is important to \textit{distribute} the counting law. However, to reproduce the angular lifespan of individual glints (to account for the smooth transition due to the micro-roughness) or their footprint, we must \textit{interpolate} between \textit{realizations} of the counting law on the grid. For the angular behaviour, we bi-linearly blend evaluations of the same binomial law:
\begin{align}
    b_{\theta_{i}}\left({w_i \times N_i, p}\right) = \sum v_{j} \, b_{\theta_{i,j}}\left({w_i \times N_i, p}\right).
\end{align}
where $v_j$ are the bi-linear weights for the four closest texel of an hemispherical grid where $\theta_{i,j}$ are the associated random seeds. We use the orthographic projection of the half-vector onto the plane to index this grid, and subdivide it to account for the glints' micro-roughness. 

Thanks to all this, we obtain an \textit{artifact-free} and \textit{stable} shader model for glints. We found that we could further improve its performance during implementation and detail how in the next Section.

\vspace{-2pt}
\section{Implementation Details}
\label{sec:implementation}
Our method heavily relies on evaluations of the binomial law. Indeed, a naive implementation of our method would lead to $10 \times 4 \times 4 = 160$ evaluations of the binomial per pixel ($2 \times 5$ anisotropic grids with linear blending of $4$ spatial texels and $4$ angular texels each). We reduce the number of calls to the binomial law by using tetrahedral discretizations (Subsection~\ref{subsec:tetrahedral_discretization}) and a simplex spatial grid (Subsection~\ref{sec:subsec_simplex_grid}), as well as the cost of evaluating the binomial law by using a gated approximation (Subsection~\ref{sec:subsec_gated_binomial})

\vspace{-5pt}
\subsection{Optimized Anisotropic Parameterization Blending}
\label{subsec:tetrahedral_discretization}

\begin{figure}[th!]
    \vspace{-0.5cm}
    \tdplotsetmaincoords{220}{70}
    \centering
    \begin{tikzpicture}[tdplot_main_coords,font=\footnotesize, scale=0.8]
    \begin{scope}[canvas is xy plane at z=-1]
        \begin{scope}
            \clip (0, 0) rectangle (5,4.2);
            
            \draw[color=blue] (0, 0) circle (1);
            \draw[color=blue] (0, 0) circle (2);
    
            \draw[color=blue, rotate=-45.0] (0, 1) -- (0, 4);
            
            \draw[color=blue] (0, 0) circle (3);
            \draw[color=blue, rotate=-22.5] (0, 2) -- (0, 4);
            \draw[color=blue, rotate=-67.5] (0, 2) -- (0, 4);
            
            \draw[color=blue] (0, 0) circle (4);
            \draw[color=blue, rotate=-11.25] (0, 3) -- (0, 4);
            \draw[color=blue, rotate=-33.75] (0, 3) -- (0, 4);
            \draw[color=blue, rotate=-56.25] (0, 3) -- (0, 4);
            \draw[color=blue, rotate=-78.75] (0, 3) -- (0, 4);
        \end{scope}
        \draw[color=blue,->] (0, 0) -- (0.0,4.5);
        \draw[color=blue,->] (0, 0) -- (4.5,0.0);

        \begin{scope}[rotate=-45]
            \node (A00) at (0.0,2.0)[orange, circle, fill, inner sep=1.5]{};
            \node (A10) at (0.0,3.0)[orange, circle, fill, inner sep=1.5]{};
        \end{scope}
        \begin{scope}[rotate=-67.5]
            \node (A01) at (0.0,2.0)[orange, circle, fill, inner sep=1.5]{};
            \node (A12) at (0.0,3.0)[orange, circle, fill, inner sep=1.5]{};
        \end{scope}
        \begin{scope}[rotate=-56.25]
            \node (A11) at (0.0,3.0)[orange, circle, fill, inner sep=1.5]{};
        \end{scope}
        \draw[thick,orange] ([shift=(45:2cm)] 0,0) arc (45:22.5:2.0cm);
        \draw[thick,orange] ([shift=(45:3cm)] 0,0) arc (45:22.5:3.0cm);
        \draw[orange,thick] (A00) -- (A10);
        \draw[orange,thick] (A00) -- (A11);
        \draw[orange,thick] (A01) -- (A11);
        \draw[orange,thick] (A01) -- (A12); 
        
    \end{scope}

    \begin{scope}[canvas is xy plane at z=0]
        \begin{scope}[opacity=0.1]
            \clip (0, 0) rectangle (5,4.2);
            
            \draw[color=blue] (0, 0) circle (1);
            \draw[color=blue] (0, 0) circle (2);
    
            \draw[color=blue, rotate=-45.0] (0, 1) -- (0, 4);
            
            \draw[color=blue] (0, 0) circle (3);
            \draw[color=blue, rotate=-22.5] (0, 2) -- (0, 4);
            \draw[color=blue, rotate=-67.5] (0, 2) -- (0, 4);
            
            \draw[color=blue] (0, 0) circle (4);
            \draw[color=blue, rotate=-11.25] (0, 3) -- (0, 4);
            \draw[color=blue, rotate=-33.75] (0, 3) -- (0, 4);
            \draw[color=blue, rotate=-56.25] (0, 3) -- (0, 4);
            \draw[color=blue, rotate=-78.75] (0, 3) -- (0, 4);
        \end{scope}
        \draw[color=blue,->, opacity=0.1] (0, 0) -- (0.0,4.5);
        \draw[color=blue,->, opacity=0.1] (0, 0) -- (4.5,0.0);

        \begin{scope}[opacity=0.5]
            \begin{scope}[rotate=-45]
                \node (B00) at (0.0,2.0)[orange, circle, fill, inner sep=1.5]{};
                \node (B10) at (0.0,3.0)[orange, circle, fill, inner sep=1.5]{};
            \end{scope}
            \begin{scope}[rotate=-67.5]
                \node (B01) at (0.0,2.0)[orange, circle, fill, inner sep=1.5]{};
                \node (B12) at (0.0,3.0)[orange, circle, fill, inner sep=1.5]{};
            \end{scope}
            \begin{scope}[rotate=-56.25]
                \node (B11) at (0.0,3.0)[orange, circle, fill, inner sep=1.5]{};
            \end{scope}
            \draw[thick,orange] ([shift=(45:2cm)] 0,0) arc (45:22.5:2.0cm);
            \draw[thick,orange] ([shift=(45:3cm)] 0,0) arc (45:22.5:3.0cm);
            \draw[orange,thick] (B00) -- (B10);
            \draw[orange,thick] (B00) -- (B11);
            \draw[orange,thick] (B01) -- (B11);
            \draw[orange,thick] (B01) -- (B12); 
        \end{scope}
    \end{scope}

    \draw[color=blue,->] (0, 0, 0) -- (0,0,-2) node[rotate=90, yshift=7pt, xshift=-0.5cm] {$s$ increase};

    \draw[orange,thick, opacity=0.5] (A00) -- (B00);
    \draw[orange,thick, opacity=0.5] (A01) -- (B01);
    \draw[orange,thick, opacity=0.5] (A10) -- (B10);
    \draw[orange,thick, opacity=0.5] (A11) -- (B11);
    \draw[orange,thick, opacity=0.5] (A12) -- (B12);
    
    \end{tikzpicture}
    \vspace{-25pt}
    \caption{
        Each cell of the anisotropic grid can be split into nine different tetrahedra. Using this tessellation, we reduce the number of binomial evaluation and improve our shader's performance.
        \label{fig:aniso_tetrahedra}
        \vspace{-15pt}
    }
\end{figure}
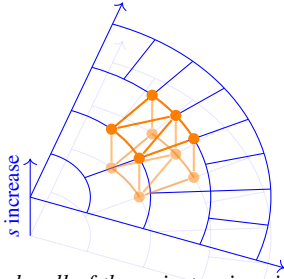

As presented in Section~\ref{sec:grids}, our anisotropic grids interpolation scheme uses a 3D grid. Since we increase the discretization of orientation at each level of the anisotropy ratio, each cell of the parameterization comprise 3 different orientations and requires $10$ vertices for distribution of the binomial law ($5$ for ratio and orientation, times $2$ for scale). We reduce this cost by tessellating this grid using tetrahedra (see Figure~\ref{fig:aniso_tetrahedra}).

In the $r-\phi$ dimension, each cell is a pentagon. We split these into three triangles (see Figure~\ref{fig:aniso_tetrahedra}~(left)). Along the $s$ dimension, each triangle becomes a prism that we can further split into three tetrahedra. The overall grid cell, can thus be decomposed into nine different tetrahedra. At run-time, we first identify which tetrahedron encompasses the pixel filter and use its barycentric coordinates to distribute the binomial law. With this, we reduce the number of binomial evaluation from $160$ down to $4 \times 4 \times 4 = 64$ per pixel.

\subsection{Simplex Surface Space Grids}
\label{sec:subsec_simplex_grid}
\begin{figure}[h!]
    \centering
    
    \def \ymax {6}
    \def \xmax {6}
    \newcommand{\myshear}[2]{(#1-0.5*#2, #2)}

    \begin{tikzpicture}[font=\footnotesize]
    \begin{scope}
        \draw (0,0) rectangle (4,3);
        \clip (0,0) rectangle (4,3);
            
        \begin{scope}[shift={(3/8+1.5, -9/8+1.5)}]
            \begin{scope}[rotate around={30:(3.0/8.0, 9.0/8.0)}]
                \begin{scope}[xscale=2]
                    \draw[step=1.0, color=green!50!black, thick, opacity=0.5]  (-3,-3) grid (6,6);   
                \end{scope}
            \end{scope}
        \end{scope}
        
    \end{scope}
    \begin{scope}[xshift=4.2cm]
        \draw (0,0) rectangle (4,3);
        \clip (0,0) rectangle (4,3);
            
        \begin{scope}[shift={(-1/8+1.5, 1.5)}]
            \begin{scope}[rotate around={30:(3.0/8.0, 9.0/8.0)}]
                \begin{scope}[xscale=2]
                    \foreach \x in {-5,...,5} {
                      \draw[gray] \myshear{\x}{-\ymax} -- \myshear{\x}{\ymax} ;
                      \draw[gray] \myshear{\x-\ymax}{-\ymax} -- \myshear{\x+\ymax}{\ymax} ;
                      \draw[gray] \myshear{-\xmax}{\x} -- \myshear{\xmax}{\x} ;
                    }
                \end{scope}
            \end{scope}
        \end{scope}
        
    \end{scope}
    \end{tikzpicture}
    \vspace{-5pt}
    \caption
    {
        \textbf{Surface Simplex.} We parameterize the surface of anisotropic grids (in green) using a simplex grid (in gray). It allows reducing the number of spatial samples from $4$ to $3$ per pixel.
        \label{fig:surface_simplex}
        \vspace{-5pt}
    }
\end{figure}
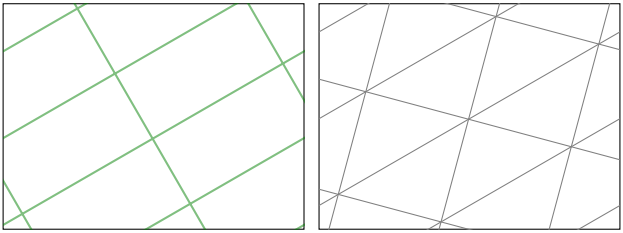

We can further decrease the amount of binomial samples by applying the same logic to the anisotropic grids themselves: instead of looking at grids with $4$ vertices per cell, we tessellate these grids into simplex grids (grids of triangles). While sampling the surface parameterization, we discretize the surface space into a 2D simplex grid instead of a regular square grid (Figure~\ref{fig:surface_simplex}) following the simplex noise method~\cite{perlin2001noise}.

Evaluation of a pixel in a triangular grid requires $3$ evaluations with barycentric interpolation, instead of $4$ bi-linearly interpolated samples. This also modifies the appearance of the glints: square grids with bi-linear interpolation produce square glints, but triangular grids with barycentric interpolation produce hexagonal glints. This results in a rounder appearance which we believe is more natural.

With this, we bring the total down to $4 \times 3 \times 4 = 48$ binomial evaluations per pixel. While we could also reduce the complexity by using a simplex grid in the angular domain, we found that the overall gain was minimal due to a bigger complexity of handling a simplex grid in this space.

\vspace{-5pt}
\subsection{Approximation of the Binomial Law}
\label{sec:subsec_gated_binomial}
\input{figures/fig_blending_with_approx}
With this many evaluations still required per pixel, the performance of sampling the binomial law becomes critical. We found that the approximation of Zirr and Kaplanyan~\cite{zirr2016real} was too costly for our use case.

\paragraph*{Classical Binomial Approximation.}
Indeed, Zirr and Kaplanyan use a Gaussian approximation of the binomial law using its first two moments (mean and variance) for which analytical forms are known: $\mathbb{E}[b(N,p)] = N \times p$ and $\mbox{Var}[b(N,p)] = N \times p \times (1-p)$.

However, to better approximate the discrete nature of the binomial at low number of trials $N$ or probability $p$, they opt to manually evaluate the sum of the Bernoulli trials:
\begin{align}
    b(N, p) \; \simeq& \; \lfloor \mathcal{N}\left(\theta_0 ; \, \mu,\, \sigma^2 \right) \rfloor & \mbox{if} \; N \times p > 5, &  \nonumber\\
    & \; \sum_{k=1}^N H(\theta_k - p) & \mbox{else}. &
\end{align}
Here, $H(x)$ is the Heaviside function. This form impacts the stability of the performance due to the varying sum length. To remedy this, we introduce a fast binomial approximation that always uses the same form, independent of the parameters $N$ and $p$.

\vspace{-3pt}
\paragraph*{Gated Gaussian Binomial Approximation.}
We devise a different approximation to the binomial distribution which only requires two random numbers to sample and no looping trials. Our approximation use the Gaussian approximation of the binomial law but aims at improving its statistics for low $N$ and $p$ parameters through the use of a single additional \emph{gating} Bernoulli trial.

We first compute the probability of having at least one success in a binomial distribution $b(N, p)$:
\begin{align}
    P_{\geq1} = 1 - (1 - p)^{N},
\end{align}
and compute the parameters $\mu, \sigma$ of a Gaussian approximation $\mathcal{N}$ of the binomial distribution for one less sample, $b(N-1, p)$:
\begin{align}
    \mu = (N-1) \times p, \;\;\; \sigma = \sqrt{(N-1) \times p \times (1 - p)}.
\end{align}
Finally, we fetch two uniform random numbers $\theta_0, \theta_1$ and discretize the continuous Gaussian distribution $\mathcal{N}$ by \emph{gating} it using a single Bernoulli trial with $p = P_{\geq1}$:
\begin{align}
    b(N, p) \approx \lfloor H(\theta_0 - P_{\geq1}) \times (1 + \mathcal{N}(\theta_1, \mu, \sigma)) \rfloor.
\end{align}
We validated that this approximation follows the reference binomial distribution (see Figure~\ref{fig:blending_with_approx}), even while blending using our distributed binomial law method, as described in Section~\ref{sec:binomials}.

As this approximation only requires two random numbers to sample, we can further speed up run-time performance by fetching them from a small texture of random numbers that we precompute once at startup, in a few milliseconds on the GPU.

\section{Results}
\label{sec:results}
We implemented our method in the Unity3D game engine, along with the methods of Zirr and Kaplanyan~\cite{zirr2016real} and of Chermain et al.~\cite{chermain2020procedural} using their provided source code. We implemented our method in both Unity's \emph{High Definition Render Pipeline} to render Figure~\ref{fig:teaser}, and Unity's built-in Forward renderer for every other Figure and for performance measurements. There, we removed any texture and disabled post-processing (except for a \emph{Bloom} filter to visualize high glint intensities). Both implementations are provided in our supplemental material and showcased in our video.

\subsection{Performance} 
We benchmarked all three methods in different rendering scenarios: the \textsc{Plane} scene where a plane is uniformly filling the screen with glinty material; the \textsc{Heightmap} scene where the glinty plane is displaced with a high-frequency heightmap; the \textsc{Sponza} scene with all objects glinty (shown in Figure~\ref{fig:results_aniso_comparison}); the \textsc{Sphere} and \textsc{Monkey} scenes with objects in the middle of the screen, as shown in Figures~\ref{fig:chermain_comparison} and~\ref{fig:ndf_convergence}. To measure the impact of anisotropy, we report different viewing configurations (facing at $90^\circ$ and grazing at $25^\circ$) for the \textsc{Plane} and \text{Heightmap} scenes. For all scenes, we use the same Beckmann NDF with a roughness of $\alpha=1.0$. To test the impact of the binomial/filtering evaluation on performances, we measured timings while varying the \textit{microfacet density} for all methods and report in Table~\ref{fig:timings} the minimum and maximum timings that we measured, on an NVIDIA RTX 2080 GPU with a $1920\times1080$ image resolution.

\begin{table}[h!]
    \centering 
    { \footnotesize \setlength{\tabcolsep}{0.6em}
    \begin{tabular}{ |c||c|c|c|c|  }    
        \hline Scene & No Glints & Ours & \cite{zirr2016real} & \cite{chermain2020procedural} \\
        \hline \textsc{Sphere}                   & 0.49 & \textbf{0.79} - \textbf{0.82} & 1.15 - 1.41 & 1.19 - 1.27 \\
        \hline \textsc{Monkey}                   & 0.49 & \textbf{0.67} - \textbf{0.68} & 0.93 - 1.09 & 1.01 - 1.04 \\
        \hline \textsc{Plane} $90^{\circ}$       & 0.49 & \textbf{1.15} - \textbf{1.25} & 1.40 - 1.83 & 1.36 - 1.70 \\
        \hline \textsc{Plane} $25^{\circ}$       & 0.49 & \textbf{1.15} - \textbf{1.26} & 2.50 - 3.18 & 3.25 - 4.61 \\
        \hline \textsc{Heightmap} $90^{\circ}$   & 0.51 & \textbf{1.52} - \textbf{1.63} & 1.93 - 2.48 & 2.19 - 2.28 \\
        \hline \textsc{Heightmap} $25^{\circ}$   & 0.51 & \textbf{1.44} - \textbf{1.54} & 3.29 - 4.15 & 4.27 - 4.83 \\
        \hline \textsc{Sponza}                   & 0.53 & \textbf{1.32} - \textbf{1.38} & 3.84 - 5.03 & 5.06 - 5.90 \\
        \hline 
    \end{tabular}
    }
    \vspace{-5pt}
    \caption
    {
        Rendering times in milliseconds per frame in our minimalist test setup. We compare the performance of a baseline Beckmann NDF function, our glinty NDF, and state-of-the-art methods~\cite{zirr2016real,chermain2020procedural}. For each test, we report the min and max timings observed while varying the microfacet density parameter. 
        \label{fig:timings}
    }
\end{table}


\paragraph*{Analysis.}
Two different aspects of our method allow it to achieve much more stable performance than other methods across the range of rendering scenarios: \textit{our anisotropic grids allow constant time evaluation} for all possible pixel footprints. Both Zirr and Kaplanyan~\cite{zirr2016real} and Chermain et al.~\cite{chermain2020procedural} rely on manually sampling the anisotropic footprint, with a varying impact on performance for each pixel. 

Additionally, \textit{our binomial approximation allows constant time evaluation} for all possible binomial parameters $N$ and $p$. Zirr and Kaplanyan~\cite{zirr2016real} rely on manually looping Bernoulli trials (up to $N = 16$ in their implementation) resulting in a variable evaluation cost, peaking right before switching to a cheaper Gaussian approximation.

The performance hit related to both aspects are further emphasized by the SIMD architecture of GPUs. With threads running in groups (\emph{waves} or \emph{warps}), a single pixel with large anisotropy (or large amount of Bernoulli trials) stalls the entire thread group, spreading the worst case scenario performance across many pixels. The impact of this can be seen in the performance hit between our \textsc{Plane} and \textsc{Heightmap} scenarios at a grazing angle, with the former concentrating the large anisotropy zones together at the top of the screen and the latter scattering pixels of large anisotropy across the entire screen.

The stable performance provided by our method is a highly desirable property for use in real-time production environments. In practice, the performance to consider for a real-time rendering method is its worst-case performance, as interactive applications cannot afford potential slowdowns in their tight frametime constraints. Furthermore, in our testing, we found that the more \textit{real use case} scenario of the \textsc{Sponza} scene, displaying a wide variety of viewing configurations across all pixels, provided the worst-case scenario and thus showed the most benefit from using our method.

\subsection{Renderings}
\begin{figure}[h!]
    \begin{tikzpicture}
        \node (Ours) { \fbox{\includegraphics[width=0.3\linewidth]{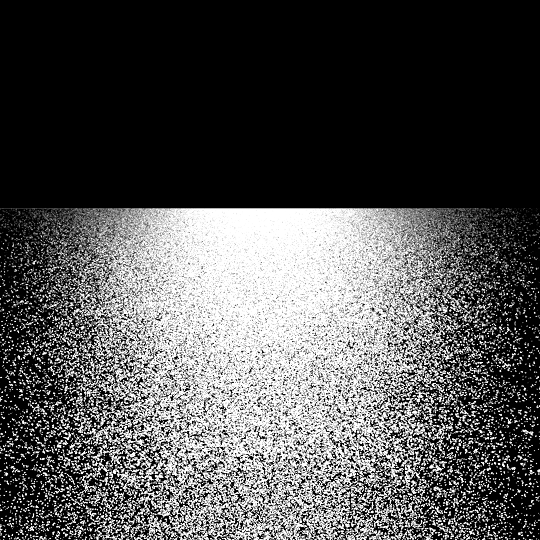}} };
        \node[below=-2pt of Ours] (tO) { a) Ours };
        
        \node[right=0pt of Ours] (Zirr) { \fbox{\includegraphics[width=0.3
        \linewidth]{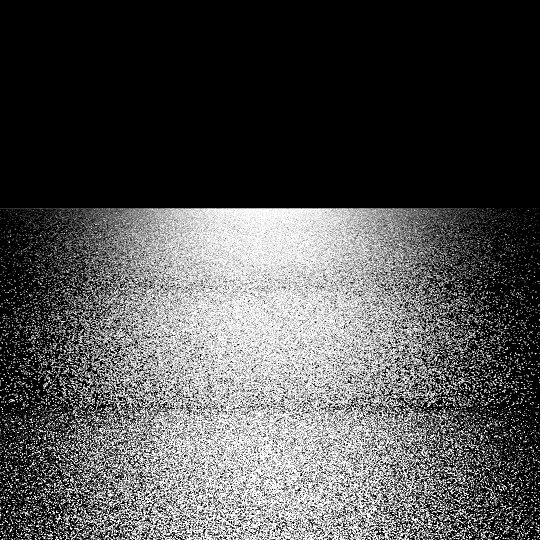}}};
        \node[below=-2pt of Zirr] (tZ) { b) {\protect\cite{zirr2016real}} };
        
        \node[right=0pt of Zirr] (Chermain) { \fbox{\includegraphics[width=0.3\linewidth]{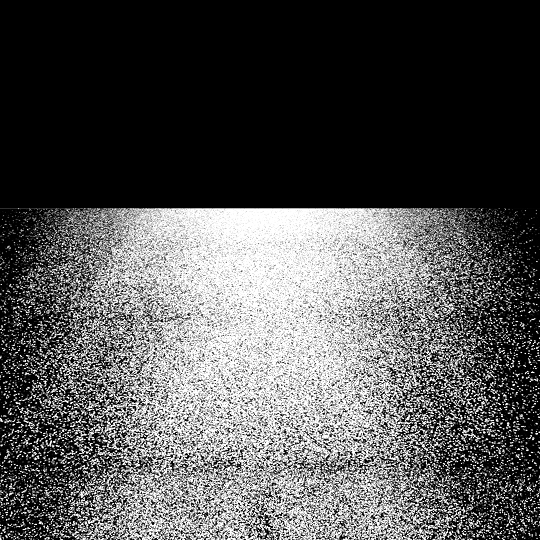}} };
        \node[below=-2pt of Chermain] (tC) { c) {\protect\cite{chermain2020procedural}} };
    \end{tikzpicture}
    \vspace{-5pt}
    \caption{
        \textbf{Blending artifacts.} Our method (a) does not suffer from blending artifacts that one can see on previous works (b-c). They experience such artifact due to linear blending between LOD levels. Such blending does not preserve the counting of glints.
        \label{fig:artefact_blend}
        \vspace{-10pt}
    }
\end{figure}

\paragraph*{Blending Artifacts.}
Previous methods~\cite{zirr2016real,chermain2020procedural} display ghosting artifacts (see Figure~\ref{fig:artefact_blend}). Those are due to incorrectly blending between either counting laws or discrete NDFs. Our glinty NDF generated using the distributed binomial laws introduced in Section~\ref{sec:binomials} doesn't suffer from this issue.

\paragraph*{Anisotropic Footprints.}
In Figure~\ref{fig:results_aniso_comparison}, we show a view point of the \textsc{Sponza} scene that results in pixel footprints with anisotropy ratios higher than $\times16$ (white area). With the maximum allowed anisotropy set to $\times16$ in both \cite{zirr2016real} and \cite{chermain2020procedural}, their evaluation cannot account for higher ratio and, hence, it implicitly increases the size of the footprint isotropically. This results in glints becoming larger on the screen in an unrealistic manner. Increasing the maximum allowed anisotropy would come at an exponentially increasing performance cost and timings variability. Using the anisotropic grid parameterizations introduced in Section~\ref{sec:grids} avoids these issues by directly sampling the surface using an appropriate parameterization.

\input{figures/fig_results_aniso_2}
\input{figures/fig_comparison_chermain}
\input{figures/fig_ndf_convergence}
\pagebreak

\paragraph*{Glinty Appearance.}
In Figure~\ref{fig:chermain_comparison}, we compare our method with the results of Chermain et al.~\cite{chermain2020procedural}. As they produce a \emph{physically-based} BRDF model that is both energy conserving, multi-scale, and continuous, we treat it as the visual reference for our method here.

We use the default parameters for the precomputed dictionary of NDFs in \cite{chermain2020procedural} and use $R = 0.034$ for our method as it closely matches it. We then vary the target Beckmann macro-scale roughness and microfacet density parameters, and capture the result of both methods with matching densities.

Our results show that we can perfectly emulate the physically-based glinty NDF in fixed images using our method. But glints feature very important temporal properties, such as smooth transitions when the viewing and/or lighting angles change. We refer to our supplemental video for temporal comparisons.

\paragraph*{Convergence to the target NDF.}

Both \cite{zirr2016real} and \cite{chermain2020procedural} limit themselves to simulating glinty NDFs converging to the Beckmann NDF, the former due to requiring analytical convolution of the NDF, the latter due to needing a separable NDF to reconstruct 2D NDFs at run-time from stored 1D NDFs.

We do not share those limitations: the formulation of our NDF as a stochastic counting process shown in Section~\ref{sec:overview} trivially converges to the target NDF $D(\mathbf{h})$. Intuitively, our formulation is solely redistributing an input energy stochastically, regardless of the function used to compute it. Our method can thus be used to achieve glinty appearances converging to any desired NDF function.

We illustrate this by targeting both the Beckmann and GGX distributions, in Figure~\ref{fig:ndf_convergence}, across varying microfacet densities. The GGX distribution has become ubiquitous in the real-time rendering industry and is the standard used in modern rendering pipeline materials, such as Unity's \emph{High Definition Render Pipeline} that we use to render the scene shown in Figure~\ref{fig:teaser}. Hence, compatibility with GGX will help the wider adoption of real-time glinty materials.

\section{Limitations \& Discussion} \label{sec:discussion}
While our model can be a direct replacement of previous works for the rendering of glints in real time, there are some cases where alternative methods would be preferable or more research is required.

\paragraph*{Anisotropic glints.}
While nothing prevents us from using an anisotropic NDF $D(\mathbf{h})$, we did not tackle the problem of rendering \textit{anisotropic glints}. Previous works link the NDF anisotropy to the micro-roughness's anisotropy as it simulates brushed metal. They do so by stretching the UV coordinates (similar to our anisotropic grids). Adding such effect in our work would create with a conflict between the footprint's anisotropy and the glint's anisotropy. For such a case, we argue that a scratch model would be more appropriate~\cite{raymond2016multi}.

\paragraph*{Grids rely on UV Mapping.}
Similar to other methods, our model requires a UV layout for both the definition of the pixel footprint (via screen-space derivatives) and the evaluation of the anisotropic grid. Unfortunately, the visual quality of the glittery appearance depends on the quality of the UV mapping. A mapping with many discontinuities will result in visible artifacts.

\paragraph*{Non-physically-based.}
Our model to render glint is not physically-based and would, for example, not pass the white furnace test. We stochastically distribute the evaluation of the NDF with potentially non-energy-conserving pixels. As shown in our results, it is not a problem in practice: we obtain renders similar to the PBR model of Chermain et al.~\cite{chermain2020procedural} in a procedural way. On the one hand, this frees us from some limitations introduced by their precomputed NDF dictionary (no fewer than four facets at once, micro-roughness constrained by macro-scale roughness). On the other hand, since we do not generate an explicit glinty NDF, we cannot importance sample it. Hence, our solution is not applicable for path tracing for example.

\paragraph*{Area lighting and Image Based Lighting.}
In Section~\ref{sec:overview} we have defined the counting process w.r.t. the NDF. Unfortunately, this does not account for the light source. As a result, we can only render punctual light with our method. Handling extended light sources would require including the integration with the light source inside the counting process.

\paragraph*{Cost per light.}
Since the counting process depends on the half-vector, our shader needs to be evaluated separately for each light. We cannot share the evaluations of the binomial law between lights. Hence, our shader cost is linear w.r.t. the number of lights.

\noindent
Both restrictions (extended lights and per light evaluation) are, to us, the last major hurdles preventing the use of our method in production. While some approximations can be done, for example distributing the result of the light integral instead of the NDF, we leave them for future work.

\section{Conclusion}
We have presented a new method to render glittery appearances in real-time. Thanks to a correct handling of the statistical counting of glints in pixel footprints and a new grid mechanism emulating anisotropic filtering, we achieve \textit{stable} performances and improve by up to $5 \times $ the performance of previous work while reducing the number of visual artifacts. While our shader is still costly compared to smooth PBR ones, we believe it will help the adoption of glittery appearances in real-time graphics such as video-games.

\bibliographystyle{eg-alpha-doi} 
\bibliography{bibliography}       

\end{document}